\documentclass{iau}
\usepackage{graphicx}

\title[FORWARD and CoMP] 
{Data-model comparison using FORWARD and CoMP}

\author[Sarah Gibson]   
{Sarah Gibson }

\affiliation{High Altitude Observatory/National Center for Atmospheric Research \\ 3080 Center Green Dr.
Boulder, CO, 80027, USA \\ email: {\tt sgibson@ucar.edu} 
}

\pubyear{2015}
\volume{305}  
\pagerange{}
\jname{Polarimetry: From the Sun to Stars and Stellar Environments}
\editors{K. N. Nagendra, S. Bagnulo, \\ R. Centeno \& M.J. Martinez Gonzalez, eds.}
\begin{document}

\maketitle

\begin{abstract}
The FORWARD SolarSoft IDL package is a community resource for model-data comparison, with a particular emphasis on analyzing coronal magnetic fields. FORWARD allows the synthesis of coronal polarimetric signals at visible, infrared, and radio frequencies, and will soon be augmented for ultraviolet polarimetry.  In this paper we focus on  observations of the infrared (IR) forbidden lines of Fe XIII, and describe how FORWARD may be used to directly access these data from the Mauna Loa Solar Observatory Coronal Multi-channel Polarimeter (MLSO/CoMP), to put them in the context of other space- and ground-based observations, and to compare them to  synthetic observables generated from magnetohydrodynamic (MHD) models.   
\end{abstract}

\firstsection 

\section{Introduction}

Given a coronal model distribution of density, temperature, velocity, and vector magnetic field, many different synthetic observables may be produced via integration along a line of sight defined by the viewer's position in heliographic coordinates.  This is the purpose of the FORWARD SolarSoft IDL package: {\it http://www.hao.ucar.edu/FORWARD/}.

FORWARD  includes routines to reproduce data from EUV/Xray imagers, UV/EUV spectrometers, white-light coronagraphs, and radio telescopes.   In addition, FORWARD links to the Coronal Line Emission (CLE) polarimetry synthesis code  (\cite[Judge \& Casini 2001]{judgecasini_01}) for forbidden coronal lines, allowing synthesis of polarimetric observations from visible to IR.

FORWARD includes several analytic magnetostatic and MHD models (\cite[Low \& Hundhausen 1995]{lowhund}, \cite[Lites \& Low 1997]{liteslow_97}, \cite[Gibson \& Low 1998]{giblow_98}, \cite[Gibson et al. 2010]{gibson_10}) in its distribution, and it is straightforward to expand it to incorporate other analytic models.  It works with user-inputted datacubes from numerical simulations, and, given calendar date input, automatically interfaces with the SolarSoft IDL Potential Field Source Surface (PFSS) package ({\it http://www.lmsal.com/$\sim$derosa/pfsspack/}) and Magnetohydrodynamics on a Sphere (MAS)-corona datacubes ({\it http://www.predsci.com/hmi/dataAccess.php}). In addition, it connects to the Virtual Solar Observatory and other web-served observations to download data in a format directly comparable to model predictions.

FORWARD creates SolarSoft IDL maps of a specified observable (e.g., Stokes I, Q, U, V).  Maps of model plasma properties (e.g., density, temperature, magnetic field, velocity) may also be created.
Plane-of-sky and Carrington plots are both standard outputs, or the user may create custom plots or simply evaluate the observable at a point or set of points in the plane of the sky. Either command-line or widget interfaces are available.  
Finally, information may be passed from one map to the next, enabling data-data, model-model, and  model-data comparison.  We now demonstrate this capability in the context of Fe XIII 1074.7 nm polarimetry.

\section{CoMP observations and data access}

The Coronal Multi-channel Polarimeter (CoMP) (\cite[Tomczyk et al. 2008]{tomczyk_08}) at the Mauna Loa Solar Observatory (MLSO) in Hawaii is comprised of a coronagraph and a narrow-band imaging polarimeter. CoMP can measure the complete Stokes I, Q, U, and V through a 0.13 nm bandpass which can be tuned across the coronal emission lines of Fe XIII at 1074.7 and 1079.8 nm and the chromospheric He\,{\sc i} line at 1083 nm.  CoMP has a 20-cm aperture and observes the full field of view of the (occulted) corona from 1.05 to 1.38 solar radii.  

CoMP Fe XIII data are available online, beginning from May 2011 and including intensity, Doppler velocity, line width, and Stokes linear polarization (Q, U as well as L = $(Q^2+U^2)^\frac{1}{2}$ and $Azimuth = 0.5*atan(\frac{U}{Q})$).  Data can be downloaded in FITS or image format via the MLSO web pages ({\it http://www2.hao.ucar.edu/mlso}).

FORWARD offers another means of downloading CoMP data, and moreover acts as a tool for the display and analysis of CoMP data.  Figure 1 demonstrates this, using the FORWARD widget interface.  In this case a calendar date has been inputted via the widget, so that the CoMP standard ``Quick Invert" data file for that date is automatically loaded when an observable is requested.  The Quick Invert file represents an averaged image and may not include all CoMP data products.  Comprehensive, non-averaged data are available in the ``Daily Dynamics" and ``Daily Polarization" FITS archives on the MLSO web pages, and once these are downloaded to a local directory they may be accessed by FORWARD by selecting ``Data: COMP: By File" in the top-left widget.

\begin{figure}[h]
\begin{center}
\includegraphics[width=5.5in]{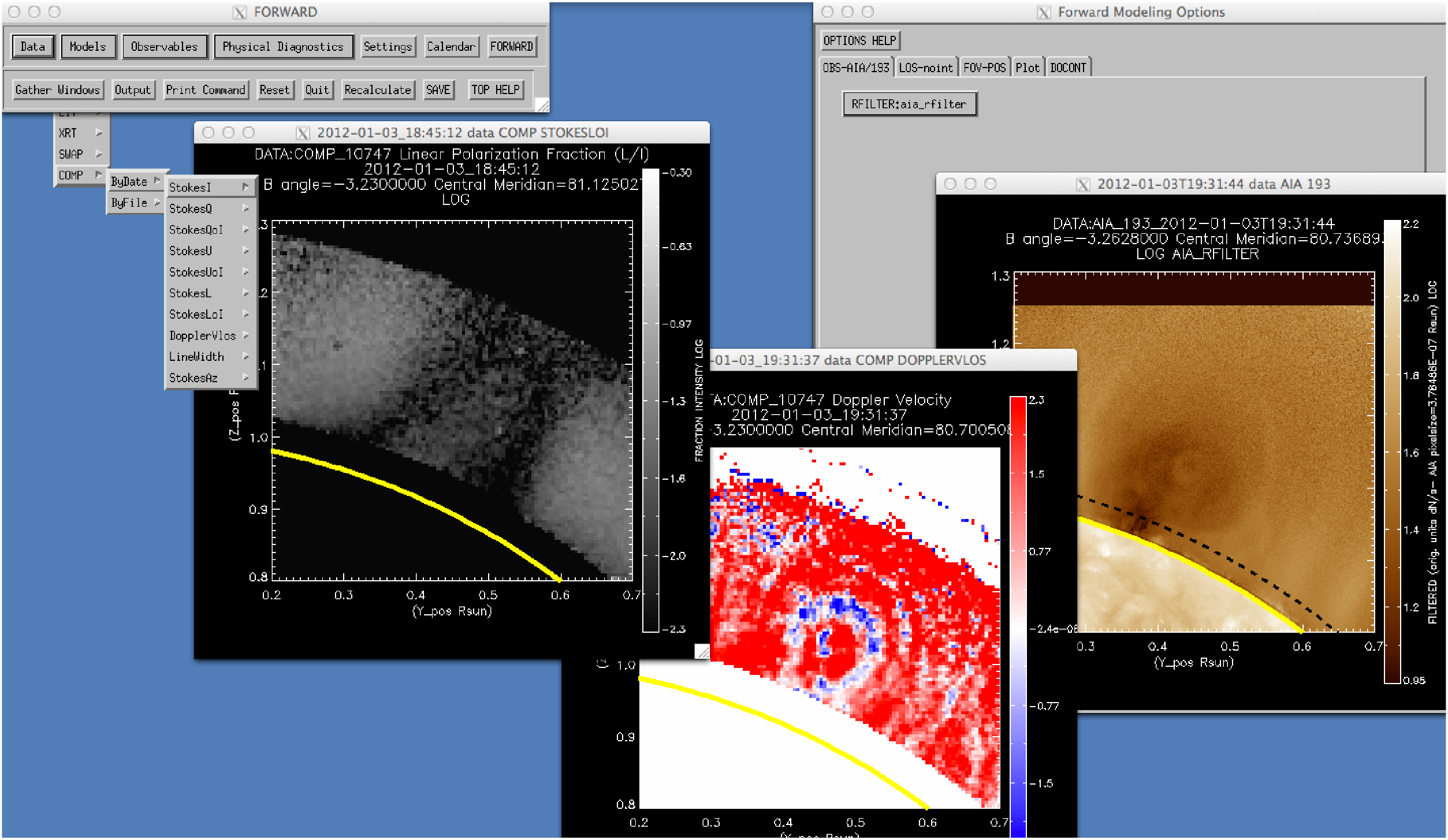} 
 \caption{Example of FORWARD codes being used to compare datasets.  When a calendar date is set (via radio button within top-left widget), data for that date can be automatically accessed as shown in the dropdown menu (top left).  Field of view and other plotting options can be set using right widget.  If keyword ``MOREPLOTS" is set (via ``Output" radio button in top-left widget) other observations can be loaded and displayed for the same date and field of view. Shown here is such a comparison for MLSO/CoMP linear polarization (left image), MLSO/CoMP Doppler velocity (middle image) and SDO/AIA 193 emission (right image).}
\end{center}
\end{figure}

At any time a SolarSoft IDL map may be saved for the data, either through the ``SAVE" radio button in top left widget (which creates an IDL save set with a unique time-stamped name), or via the ``Output" radio button which allows customized naming.  Similarly, images may be saved in TIFF or Postscript format (further image format options are available if using the line command version of FORWARD).

\section{FORWARD synthesis of CoMP-type observations}

The CLE code of \cite[Judge \& Casini (2001)]{judgecasini_01} is called by FORWARD to synthesize Stokes profiles for the Fe XIII forbidden lines.  The code assumes that the lines are optically thin, and that they are excited by both anisotropic photospheric radiation and thermal particle collisions.  Magnetism manifests in these lines firstly through circular polarization (Stokes V) arising from the longitudinal Zeeman effect, and secondly through a depolarization of the linearly-polarized line through the saturated Hanle effect.  The linearly polarized light (Q,U) has approximately 100 times more signal than the circularly polarized light (V), and so we focus on it from hereon.

\begin{figure}[h]
\begin{center}
\includegraphics[width=4.8in]{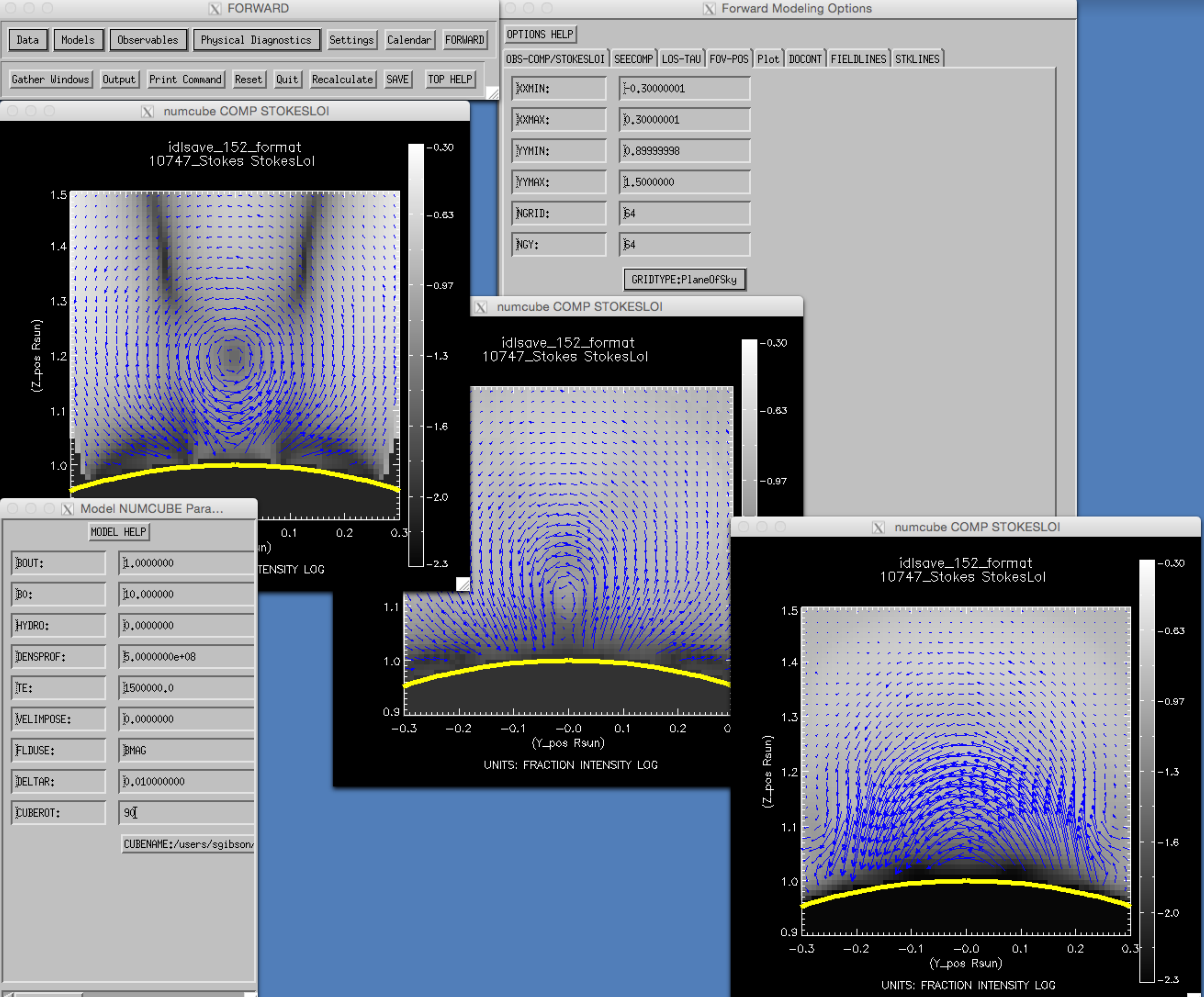} 
 \caption{Example of FORWARD codes being used to vary a model property - in this case flux-rope orientation - and to demonstrate the resulting impact on predicted linear polarization. Model parameters are controled by the bottom-left widget, in this case we vary the ``CUBEROT" parameter that changes the orientation of the numerical flux-rope data cube which we are forward modeling (note the actual line-of-sight integration is triggered by pushing the button ``FORWARD" on the top-left widget).  The top-left image is a flux rope oriented along the viewer's line of sight, and results in a lagomorphic (rabbit-head shaped) linear polarization signal. The middle image is a rope oriented 45 degrees to the line of sight, and the bottom-right image is a rope oriented perpendicular to the line of sight. The grey-scale image shows fraction of linearly polarized light (log scale), and the vectors show the direction of the model magnetic field in the plane of the sky.   }
\end{center}
\end{figure}

In order to calculate the Stokes profiles, it is necessary to specify a distribution of electron density, temperature, velocity, and vector magnetic field in the corona.  A range of such model coronae have been specified and analyzed, resulting in a growing body of literature that demonstrates the usefulness of the linear polarization of Fe XIII 1074.7 nm for identifying characteristic magnetic topologies in the corona (\cite[Judge et al. 2006]{judge_06}, \cite[Dove et al. 2011]{dove_11}, \cite[Rachmeler et al. 2012; 2013; 2014]{rachmeler_12,rachmeler_13,rachmeler_14}, \cite[{B{\c a}k-St{\c e}{\'s}licka} {et~al.} 2013; 2014]{ula_13,ula_14}, \cite[Gibson 2014; 2015]{gibsoniau_14,gibson_15}).

Figure 2 shows an example of how FORWARD can be used to vary properties of a model corona and examine how this affects predicted linear polarization.  In this case a model of a magnetic flux rope (\cite[Fan 2012]{fan_12}) predicts a ``lagomorphic" (rabbit-shaped) structure in linear polarization.   When the flux rope orientation is varied, the lagomorphic structure loses visibility.  Such linear-polarization lagomorphs are observed by CoMP (\cite[{B{\c a}k-St{\c e}{\'s}licka} {et~al.} 2013]{ula_13}) in association with polar crown filament cavities, which are oriented largely parallel to the viewer's line of sight.

\begin{figure}[h]
\begin{center}
 \includegraphics[width=5in]{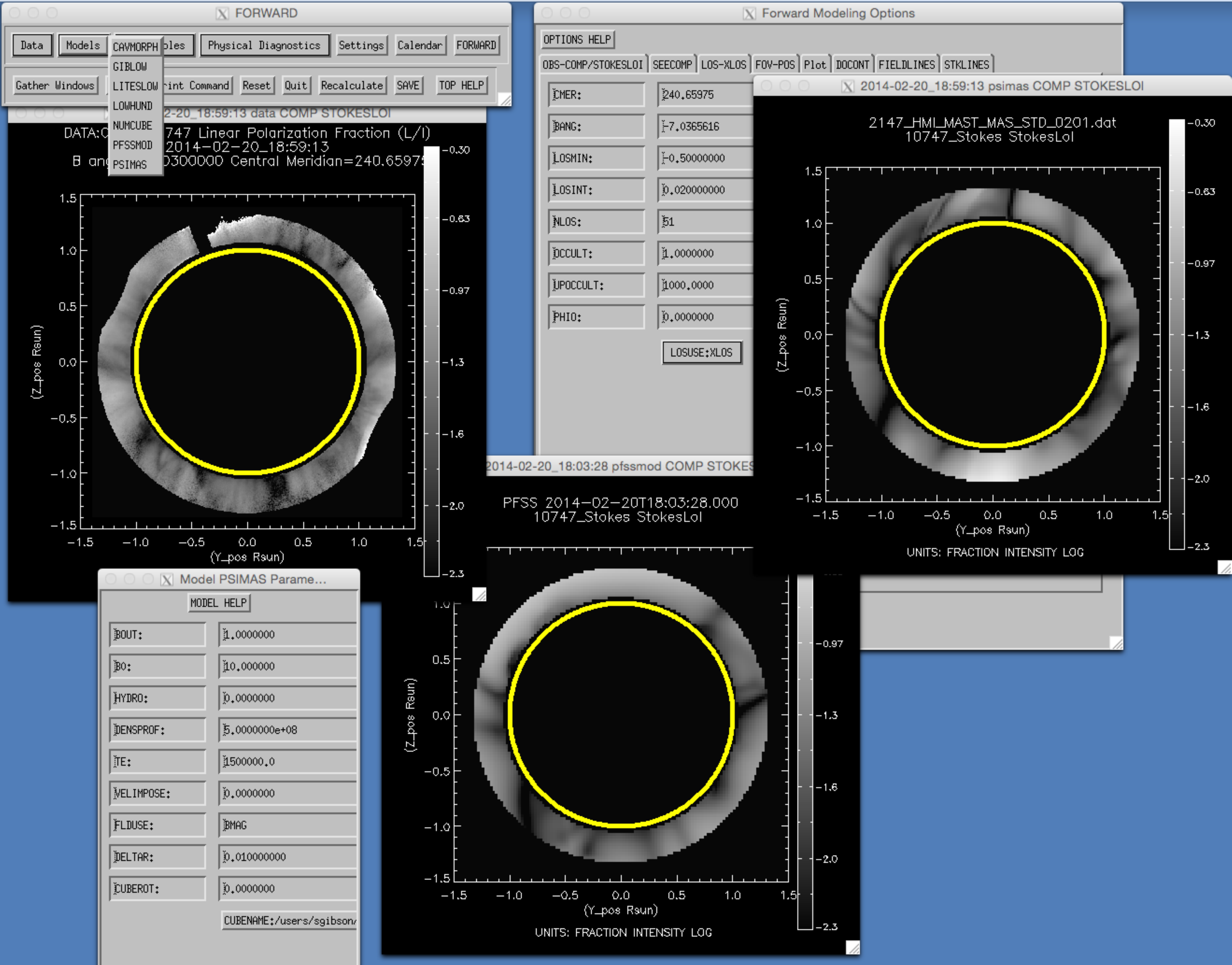} 
\caption{Example of FORWARD codes being used to synthesize linear polarization distributions from global magnetic models, and to compare these to actual CoMP observations.   In addition to providing the choice of a set of analytic models and the option of inputting one's own numerical data cube, the ``Models" radio button on the top-left widget allows access to the global PFSS and MAS models.  A date can be set using the ``Calendar" radio button and these global models automatically accessed over the internet, or alternatively the user may point to a local copy of the global model datacube using the ``CUBENAME" radio button in the bottom-left widget (note that a local copy is saved by default, and sought before going to the internet to download).  By employing the ``MOREPLOTS" keyword, we show linear polarization for CoMP data (left image), the forward-modeled PFSS (bottom middle), and the forward-modeled MAS (right top) for the same date, field of view, and viewer's position. }
\end{center}
\end{figure}

\section{Comparison of CoMP data to FORWARD-synthesized observables}

In addition to identifying the linear polarization signatures of particular types of structures such as flux ropes, FORWARD may be used to directly compare linear polarization synthesized from global magnetic models to CoMP data.  Figure 3 shows an example of such a comparison.  Some features are common to all three, such as the dark curved feature originating from approximately polar angle 130$^\circ$ (measured counter-clockwise from North).  Other features may be better captured by the MAS model than the PFSS, such as the parallel curved features near polar angle 90$^\circ$.  Still others are not captured by either model, such as the lagomorphic structures between approximately polar angle 330-350.  If the lagomorphs are polar crown filament flux ropes that are built up over days or even weeks, it is not surprising that the models -- which do not employ a time-varying photospheric magnetic field as a boundary condition -- might miss them.  We must be cautious in interpreting these differences, however, as they may arise from time evolution, model assumptions, data uncertainty, or some combination of all of these.

\section{Conclusions}

FORWARD is available as a SolarSoft package, and further documentation can be found at the web address referenced above.  It continues to grow as capabilities in new wavelength regimes are added.  For coronal magnetometry, comparing model to multi-wavelength data is particularly important, since different wavelengths probe different parts of the corona.  The goal of FORWARD is to act as a community framework to facilitate such comprehensive model-data comparisons.

\acknowledgements  
CoMP data are courtesy of the Mauna Loa Solar Observatory, operated by the High Altitude Observatory, as part of the National Center for Atmospheric Research (NCAR).  NCAR is supported by the National Science Foundation.   I thank the Air Force Office of Scientific Research for support under Grant FA9550-15-1-0030.   I thank Terry Kucera, James Dove, Laurel Rachmeler, Blake Forland, Tim Bastian, Stephen White, Silvano Fineschi, Cooper Downs, Haosheng Lin, Don Schmit, Kathy Reeves, and Yuhong Fan for their code contributions to FORWARD, as well as the other members of the International Space Science Institute (ISSI)  teams on coronal cavities (2008-2010) and coronal magnetism (2013-2014) who were key to guiding FORWARD development efforts.  In addition, I acknowledge Steve Tomczyk and Chris Bethge for assistance with interfacing FORWARD with the CoMP data, Phil Judge and Roberto Casini for assistance with interfacing FORWARD with the CLE code, Cooper Downs and Jon Linker for assistance interfacing with the PSI/MAS model,  Marc de Rosa for assistance with interfacing with his PFSS SolarSoft codes,  Dominic Zarro for assistance with VSO interfacing, and Sam Freeland for assistance with SolarSoft interfacing.



\begin{thebibliography}{13}
\expandafter\ifx\csname natexlab\endcsname\relax\def\natexlab#1{#1}\fi

\bibitem[{{B{\c a}k-St{\c e}{\'s}licka} {et~al.}(2013){B{\c a}k-St{\c
  e}{\'s}licka}, Gibson, Fan, Bethge, Forland, \& Rachmeler}]{ula_13}
{B{\c a}k-St{\c e}{\'s}licka}, U., Gibson, S.~E., Fan, Y., Bethge, C., Forland,
  B., \& Rachmeler, L.~A. 2013, {\it ApJ} 770, 28

\bibitem[{{B{\c a}k-St{\c e}{\'s}licka} {et~al.}(2014){B{\c a}k-St{\c
  e}{\'s}licka}, Gibson, Fan, Bethge, Forland, \& Rachmeler}]{ula_14}
{B{\c a}k-St{\c e}{\'s}licka}, U., Gibson, S.~E., Fan, Y., Bethge, C., Forland,
  B., \& Rachmeler, L.~A. 2014, in: B. Schmieder, J.-M. Malherbe \& 
 S.T. Wu (eds.), {\textit {Nature of Prominences and their role in Space 
Weather}}, Proc. IAU Symposium No. 300 (Cambridge: CUP), p.~395

\bibitem[{Dove {et~al.}(2011)Dove, Gibson, Rachmeler, Tomczyk, \&
  Judge}]{dove_11}
Dove, J., Gibson, S., Rachmeler, L.~A., Tomczyk, S., \& Judge, P. 2011,
  {\textit {ApJ}} 731, 1

\bibitem[{{Fan}(2012)}]{fan_12}
{Fan}, Y. 2012, {\textit {ApJ}} 758, 60

\bibitem[{{Gibson}(2014)}]{gibsoniau_14}
{Gibson}, S. 2014, in: B. Schmieder, J.-M. Malherbe \& 
 S.T. Wu (eds.), {\textit {Nature of Prominences and their role in Space 
Weather}}, Proc. IAU Symposium No. 300 (Cambridge: CUP), p.~139

\bibitem[{{Gibson}(2015)}]{gibson_15}
{Gibson}, S. 2015, in: 
 {\textit {Solar Prominences, Astrophysics and Space Science Library}}, Volume 415 (Springer), p.~323

\bibitem[{Gibson {et~al.}(2010)Gibson, Kucera, Rastawicki, Dove, de~Toma, Hao,
  Hill, Hudson, Marque, McIntosh, Rachmeler, Reeves, Schmieder, Schmit, Seaton,
  Sterling, Tripathi, Williams, \& Zhang}]{gibson_10}
Gibson, S.~E., Kucera, T.~A., Rastawicki, D., Dove, J., de~Toma, G., Hao, J.,
  Hill, S., Hudson, H.~S., Marque, C., McIntosh, P.~S., Rachmeler, L., Reeves,
  K.~K., Schmieder, B., Schmit, D.~J., Seaton, D.~B., Sterling, A.~C.,
  Tripathi, D., Williams, D.~R., \& Zhang, M. 2010, {\textit {ApJ}} 723, 1133

\bibitem[{Gibson \& Low(1998)}]{giblow_98}
Gibson, S.~E., \& Low, B.~C. 1998, {\textit {ApJ}} 493, 460

\bibitem[{{Judge} \& {Casini}(2001)}]{judgecasini_01}
{Judge}, P.~G., \& {Casini}, R. 2001, in: {M.~Sigwarth} (ed.), 
{\textit {Advanced Solar Polarimetry -- Theory,
  Observation, and Instrumentation}}, ASP 
  Conf. Series 236 (San Francisco: ASP), p. 503

\bibitem[{Judge {et al.}(2006){Judge},{Low}, \& {Casini}}]{judge_06}
Judge, P.~G., Low, B.~C., \& Casini, R. 2006, {\textit {ApJ}} 651, 1229

\bibitem[{Lites \& Low(1997)}]{liteslow_97}
Lites, B.~W., \& Low, B.~C. 1997, {\textit {Solar Phys.}} 174, 91

\bibitem[{Low \& Hundhausen(1995)}]{lowhund}
Low, B.~C., \& Hundhausen, J.~R. 1995, {\textit {ApJ}} 443, 818

\bibitem[{{Rachmeler} {et~al.}(2012){Rachmeler}, {Casini}, \&
  {Gibson}}]{rachmeler_12}
{Rachmeler}, L.~A., {Casini}, R., \& {Gibson}, S.~E. 2012, in: 
T.~R. {Rimmele}, A.~{Tritschler},
  F.~{W{\"o}ger}, M.~{Collados Vera}, H.~{Socas-Navarro}, R.~{Schlichenmaier},
  M.~{Carlsson}, T.~{Berger}, A.~{Cadavid}, P.~R. {Gilbert}, P.~R. {Goode}, \&
  M.~{Kn{\"o}lker} (eds.), {\textit  {The Second ATST-EAST Meeting: Magnetic 
Fields from the Photosphere to the Corona}}, ASP Conf. Series 463 (San 
Francisco: ASP), p.~227

\bibitem[{Rachmeler {et~al.}(2013)Rachmeler, Gibson, Dove, DeVore, \&
  Fan}]{rachmeler_13}
Rachmeler, L.~A., Gibson, S.~E., Dove, J.~B., DeVore, C.~R., \& Fan, Y. 2013,
  {\textit {ApJL}} 787, L3

\bibitem[{Rachmeler {et~al.}(2014)Rachmeler, Platten, Bethge, Seaton, \&
  Yeates}]{rachmeler_14}
Rachmeler, L.~A., Platten, S. J., Bethge, C., Seaton, D. B., \& Yeates, A. R., 2014,
  {\textit {Solar Phys.}} 288, 617 

\bibitem[{Tomczyk {et~al.}(2008)Tomczyk, Card, Darnell, Elmore, Lull, Nelson,
  Streander, Burkepile, Casini, \& Judge}]{tomczyk_08}
Tomczyk, S., Card, G.~L., Darnell, T., Elmore, D.~F., Lull, R., Nelson, P.~G.,
  Streander, K.~V., Burkepile, J., Casini, R., \& Judge, P.~G. 2008, 
 {\textit {Solar Phys.}} 247, 411

\end{thebibliography}



\end{document}